# A Requirements Engineering Technology for the IoT Software Systems

**Danyllo Valente da Silva** [Federal University of Rio de Janeiro | dvsilva@cos.ufrj.br]
**Bruno Pedraça de Souza** [Federal University of Rio de Janeiro | bpsouza@cos.ufrj.br]
**Taisa Guidini Gonçalves** [Federal University of Rio de Janeiro | taisa@cos.ufrj.br]
**Guilherme Horta Travassos** [Federal University of Rio de Janeiro | ght@cos.ufrj.br]

**Abstract**

Contemporary software systems (CSS) – such as the internet of things (IoT) based software systems – incorporate new concerns and characteristics inherent to the network, software, hardware, context awareness, interoperability, and others, compared to conventional software systems. In this sense, requirements engineering (RE) plays a fundamental role in ensuring these software systems' correct development looking for the business and end-user needs. Several software technologies supporting RE are available in the literature, but many do not cover all CSS specificities, notably those based on IoT. This research article presents $RET_{IoT}$ (Requirements Engineering Technology for the Internet of Things based software systems), aiming to provide methodological, technical, and tooling support to produce IoT software system requirements document. It is composed of an IoT scenario description technique, a checklist to verify IoT scenarios, construction processes, and templates for IoT software systems. A feasibility study was carried out in IoT system projects to observe its templates and identify improvement opportunities. The results indicate the feasibility of $RET_{IoT}$ templates' when used to capture IoT characteristics. However, further experimental studies represent research opportunities, strengthen confidence in its elements (construction process, techniques, and templates), and capture end-user perception.

***Keywords:*** *Software Engineering, Requirements Engineering, Internet of Things, IoT Software Systems, Software Systems Specification, Software system requirements document, Software technology*

## 1 Introduction

Contemporary software systems, such as those inherent to the Internet of Things (IoT) paradigm, are complex compared to conventional software systems. This complexity comes from the inclusion of new concerns and characteristics related to network, software, hardware, context awareness, interface, interoperability, and others (Motta et al., 2019a) (Nguyen-Duc et al., 2019).

IoT-based software systems seek to promote the interlacement of technologies and devices that, through a network, can capture and exchange data, make decisions, and act. With these actions, they unite the real and virtual worlds through objects and tags. Due to its specific technological characteristics, building IoT software systems is not a trivial activity, requiring adapted and/or innovative software technologies to create and guarantee the quality of the built product (Motta et al., 2019a).

The quality of contemporary software systems' development depends on software technologies that respond to these systems' new concerns and characteristics. As with any other product built on engineering principles, a key activity in developing IoT software systems is constructing the requirements document. Defects present in the requirements document can cause an increased time, cost, and effort for the project; dissatisfied customers and end-users; low reliability of the software system; a high number of failures; among others (Vegendla et al. 2018) (Arif et al. 2009).

Requirements engineering (RE) is responsible for the life cycle of the requirements document and ensures its proper construction (Vegendla et al. 2018) (Pandey et al., 2010).

The RE phases and activities may differ according to the application domain, people involved, processes, and organizational culture. However, we can observe some recurring phases and RE activities, such as conception/design, elicitation, negotiation, analysis, specification, verification, validation, and management.

The technical literature presents several software technologies to support RE for software systems, but not all of them cover the different RE phases and, mainly, IoT software systems' specificities. In this work, the term "software technology" refers to the methodological, technical, and tooling offered by the works to support the construction of requirements documents for IoT software systems.

Considering the need for appropriate software technologies for the development of IoT software systems and understanding the importance of requirements documents for the stability, adequacy, and quality of a project, this work proposes the $RET_{IoT}$ (Requirements Engineering Technology for the Internet of Things software systems).

The $RET_{IoT}$ consists of a requirements specification technique based on IoT scenarios description - $SCENARI_{OT}$ (Silva 2019), an IoT scenario inspection technique - $SCENARI_{OT}CHECK$ (Souza 2020), a construction process, and templates to support the processes activities and build the requirements document.

The $SCENARI_{OT}$ and $SCENARI_{OT}CHECK$ techniques were previously evaluated through experimental studies, which indicated their feasibility (Souza et al. 2019a) and usefulness (Souza et al. 2019b). Moreover, they have been used in IoT software system projects developed by the Experimental Software Engineering (ESE) Group in the context of DELFOS – The Observatory of Engineering Contemporary Software Systems – COPPE/UFRJ.

Based on the experiences with these projects, the construction process, and templates of RET$_{IoT}$ evolved. This article extends a previous publication of RET$_{IoT}$ (Silva et al. 2020b). The first version (Silva et al. 2019) encompasses many RE activities and focuses on the definition of Project scope, IoT system, and IoT system requirements. The second version (Silva et al. 2020b) focuses on eight RE phases: Conception, IoT elicitation, IoT analysis, IoT specification, IoT verification, Negotiation, Validation, and Management. The templates of this version are evaluated through a feasibility study (section 4). The third version (Silva et al. 2020a) involves the different RE phases through an engineering cycle divided into eight phases: IoT ideation and conception, IoT elicitation, IoT analysis, IoT specification, IoT verification, Negotiation, IoT evaluation, and Management. Their templates are evaluated through a proof of concept. The fourth and current version of the technology is composed of four phases through an engineering cycle. Also, it encompasses product ideation and evaluation concepts.

For the sake of completeness and applicability, this paper presents the current (fourth) version of RET$_{IoT}$, including a feasibility study comparing three RET$_{IoT}$ templates with regular ones used to build requirements documents for conventional software systems. The results indicate that RET$_{IoT}$ templates allow capturing the information needed for IoT software systems and that they are mature to be evaluated in constructing such software requirements documents. It is also possible to observe that the technology covers the main RE phases and activities concerning such IoT-based projects.

Beyond this introduction, this article presents six other sections. Section 2 describes the technological basis of the RET$_{IoT}$. Next, Section 3 introduces and details the RET$_{IoT}$. Section 4 demonstrates the feasibility study. Section 5 presents some related works found in the literature. Section 6 discusses some research opportunities. Finally, section 7 presents future work and concludes the article.

## 2 The Technological Basis of the RET$_{IoT}$

This section presents the technological basis used to build the RET$_{IoT}$ to support RE in IoT software systems. Such a requirements technology is inserted in the context of a systems engineering approach, which concerns the major development stages of IoT software systems (Motta et al. 2020). Its technological basis is composed of two empirically evaluated techniques, the SCENARI$_{OT}$ and SCENARI$_{OT}$CHECK techniques.

### 2.1 SCENARI$_{OT}$

Conventional software scenarios can be used in any software system and development stage, covering different purposes, such as eliciting requirements, specifying requirements, validating requirements, and testing (Glinz 2000) (Behrens 2002) (Alexander and Maiden 2004). A scenario is a sequence of events describing the system behavior and its environment (Burg and Van de Riet 1996) or an ordered set of interactions between partners - usually systems and external actors (Glinz 2000). It represents requirements through stories describing the system from the users' perspective when applied to requirements engineering (Glinz 2000) (Alexander and Maiden 2004).

Scenarios offer many advantages: they are based on the users' point of view; ii) the possibility to carry out partial specifications; iii) easy to understand; iv) enable short feedback loops; and v) provide a basis for testing the system (Glinz 2000). Thus, scenarios constitute a good basis for communication with clients and laypeople (non-technical) because they can be easily understood and do not require prior understanding. Therefore, everyone involved at different levels and functions can express opinions and identify problems (Glinz 2000) (Behrens 2002) (Alexander and Maiden 2004).

The SCENARI$_{OT}$ (Silva 2019) is a specification technique that adapts conventional scenarios to support IoT software systems' specifications. It considers the characteristics (adaptability, connectivity, privacy, intelligence, interoperability, mobility, among others) and behaviors (identification, sensing, and actuation) specific to these software systems (Motta et al., 2019b). The combination of characteristics and behaviors led to the creation of nine IoT Interaction Arrangements (IIAs).

IoT interaction arrangements represent frequent interaction flows between things and other non-IoT elements, such as conventional software systems and end-users. Each IIA has a catalog containing all relevant information captured and used in the scenario's description. The cardinality for arrangements and scenarios is a many-to-many relationship (M:N). Therefore, many arrangements (isolated or combined) relate to one or more IoT scenarios, and an IoT scenario can be linked to one or more arrangements.

The IIAs, together with their catalogs, guide software engineers to capture essential information about the system: i) identification of the "things" and system components; ii) the types of data that will be collected and displayed; iii) the actions that will be performed in the environment; iv) aspects related to decision making on a particular system context; v) the actors (end-users, software systems, things, among others) who will access the data; among others. **Figure 1** shows the "IIA-1: Display of IoT data" arrangement and its catalog.

### 2.2 SCENARI$_{OT}$CHECK

The SCENARI$_{OT}$CHECK is a checklist-based software inspection technique specialized in verifying IoT software system scenarios (Souza 2020). This technique aims to assist inspectors in detecting defects in IoT scenario descriptions, guaranteeing their quality. It was created to act together with the SCENARI$_{OT}$ technique since it produces the input to SCENARI$_{OT}$CHECK.

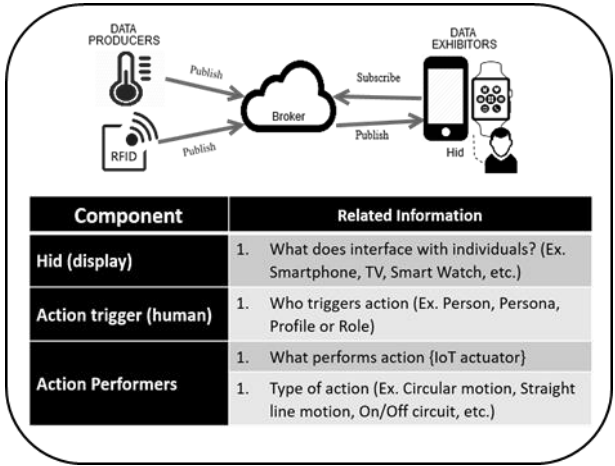

**Figure 1.** "IIA-1: Display of IoT data" arrangement (Silva 2019).

The SCENARI$_{OT}$CHECK checklist consists of two parts. The first part (general questions) aims to identify defects related to project information and systemic solution such as i) problem domain; ii) interaction and identification among actors, system, hardware, and devices; iii) alternative and exception flow; among others. The second part (specific questions) considers the non-functional properties of IoT software systems. **Figure 2** presents an overview of the checklist.

After specifying IoT scenarios, the inspectors can apply the SCENARI$_{OT}$CHECK technique to verify the scenario descriptions. The identified non-conformities are described in the inspection report. Finally, after the discrimination meeting (defects identification), the IoT scenario specification document is corrected. The application process of the two techniques is shown in **Figure 3**.

The SCENARI$_{OT}$CHECK complements SCENARI$_{OT}$ by providing a template for IoT scenarios specification. This template resembles a use-case description document with some additional fields: i) identification of the IoT software system elements; ii) problem domain description, iii) role description of each actor involved in the scenario;

and iv) interactions description between the actors (end-user, things, software system, among others) and the IoT software system.

**Figure 2.** Checklist overview.

## 3 The RET$_{IoT}$

The RET$_{IoT}$ (Requirements Engineering Technology for the Internet of Things based software systems) comprises the techniques described in section 2, a construction process, and templates to build requirements documents following RE principles.

The requirements documents' construction process is based on the main RE phases (Pressman 2014) (Sommerville 2015): conception/design, elicitation, analysis, specification, negotiation, verification, validation, and management. However, the RET$_{IoT}$ adapts and includes new activities to meet the specificities of IoT software systems.

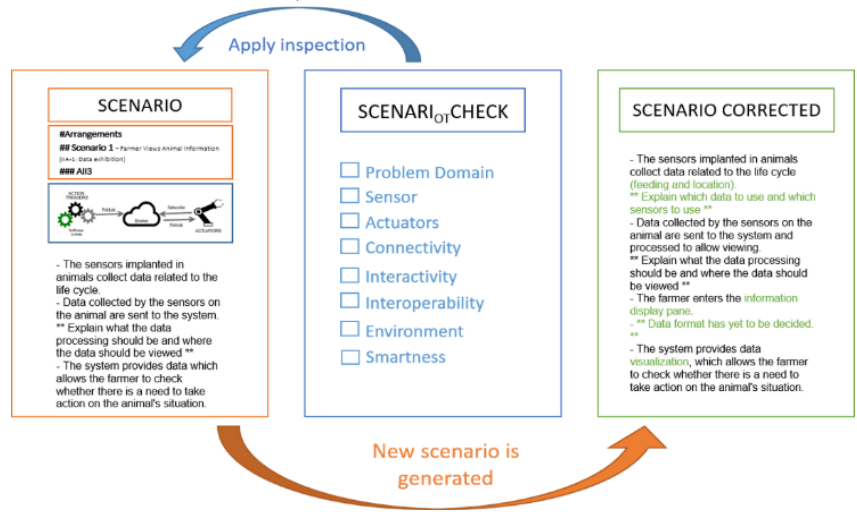

**Figure 3.** Application process of SCENARI$_{OT}$ and SCENARI$_{OT}$CHECK techniques (Souza et al. 2019a).

## 3.1 Construction process

The current version of the technology encompasses product ideation, evaluation concepts, such as low- and high-level prototypes, and MVPs' creation (Minimum Viable Product) for the desired product. The construction process incorporates aspects and characteristics found in the literature review inherent to IoT software systems.

It also involves the different RE phases (Pressman 2014) (Sommerville 2015) through an engineering cycle divided into four phases: **IoT ideation, conception, elicitation; IoT analysis and specification; IoT negotiation and evaluation;** and **Management.**

**Figure 4** presents an overview of the construction process engineering cycle with two dimensions: **main** and **transversal** (performed in parallel). The main dimension corresponds to the activities and tasks required to build the IoT requirements document.

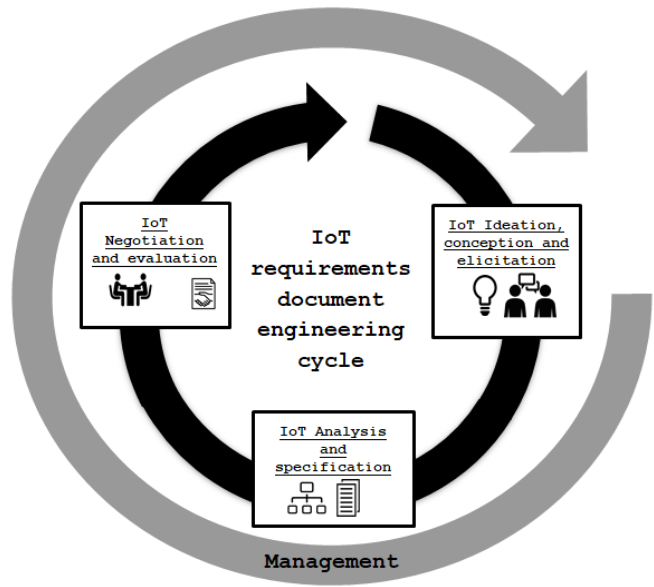

**Figure 4.** Construction process overview.

The transversal dimension (see **Figure 5**) offers three management activities and tasks focused on the artifacts and process management. The activities and tasks do not have a specific and determined time to be performed. Everything depends on the need identified by the user through the main process flow.

The technology proposes version control of the artifacts and traceability between requirements, IoT scenarios, IoT interaction arrangements, and IoT use cases in the management phase. Besides, RET$_{IoT}$ offers change management so that modifications in requirements can be reflected in the generated artifacts.

The construction of the IoT requirements document is performed iteratively and incrementally. The engineering cycle is executed three times, where each execution is called a **stage.**

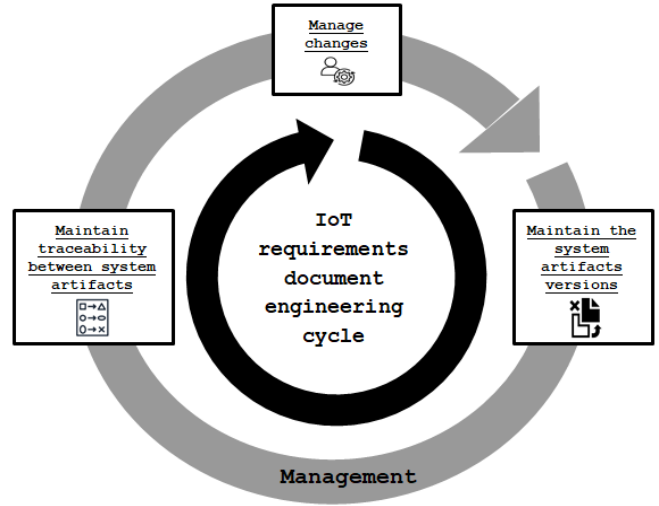

**Figure 5.** Management phase overview.

## 3.2 Construction process stages

Each stage performs the common phases (see **Figure 4** and **Figure 5**) of the engineering cycle generating intermediate artifacts. The result of Stage 3 is the IoT requirements document. However, each stage has specific objectives, activities, and tasks. The construction process can be executed for one idea or a set of requirements. In the first case, the process execution is an intermediated version of the IoT requirements document. Also, the construction process can be used with any development methodology. For example, if the project uses an agile methodology, the IoT Use cases are not mandatory.

Besides, the current RET$_{IoT}$ version integrates ten templates – eight of them are defined/adapted from the project templates currently used in projects of the ESE group/PESC/COPPE and other templates used in software engineering group/PESC/COPPE, and two of them were defined by SCENARI$_{OT}$CHECK technique (Souza 2020).

**Figure 6** shows an overview (IDEF0 diagram) of the three stages with their inputs, outputs, templates presented in the next paragraphs, and controls - Management procedures and Feasibility strategy. The Management phase performs the management procedures. The Feasibility strategy represents the milestone of each Stage.

### 3.2.1 Stage 1

The first stage is to **understand the problem.** It aims to understand the problem or opportunity, analyze the stakeholders and their needs, elicit the business needs and carry out the project feasibility analysis. It is composed of 12 activities and 27 tasks that are distributed throughout the engineering cycle. **Figure 7** presents an overview of the activities performed in the first stage. This stage offers three templates: **IoT Canvas, IoT Project Feasibility Analysis,** and **Requirements Checklist.** Its milestone is the **Feasibility Analysis** performed by four activities (Analyze market demand, Analyze economic feasibility, Analyze impact and risks, and Analyze technical feasibility).

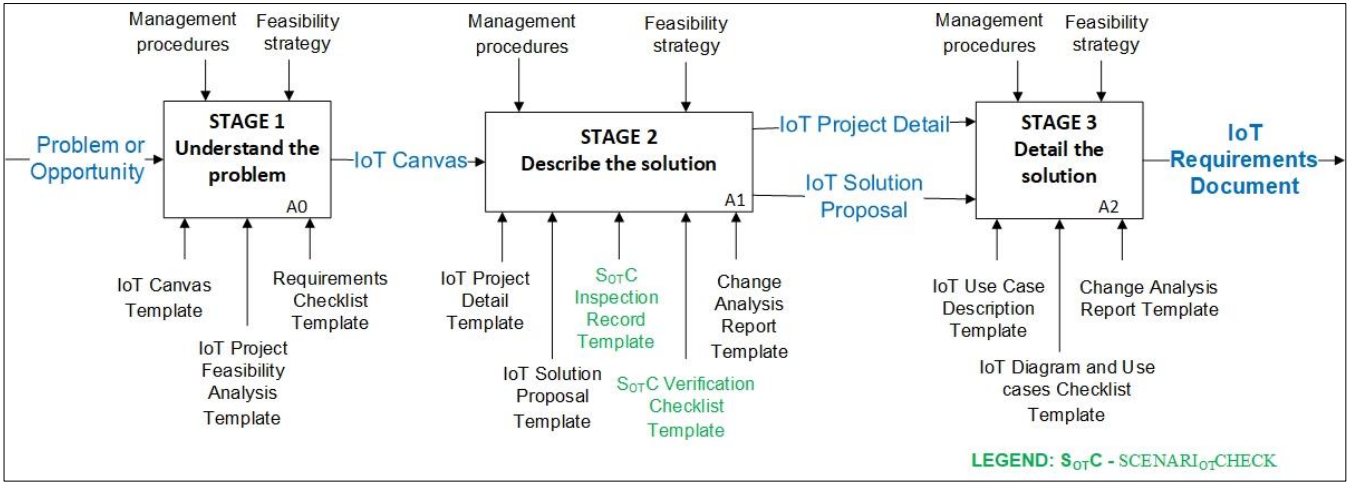

**Figure 6.** IDEF0 diagram of the three stages

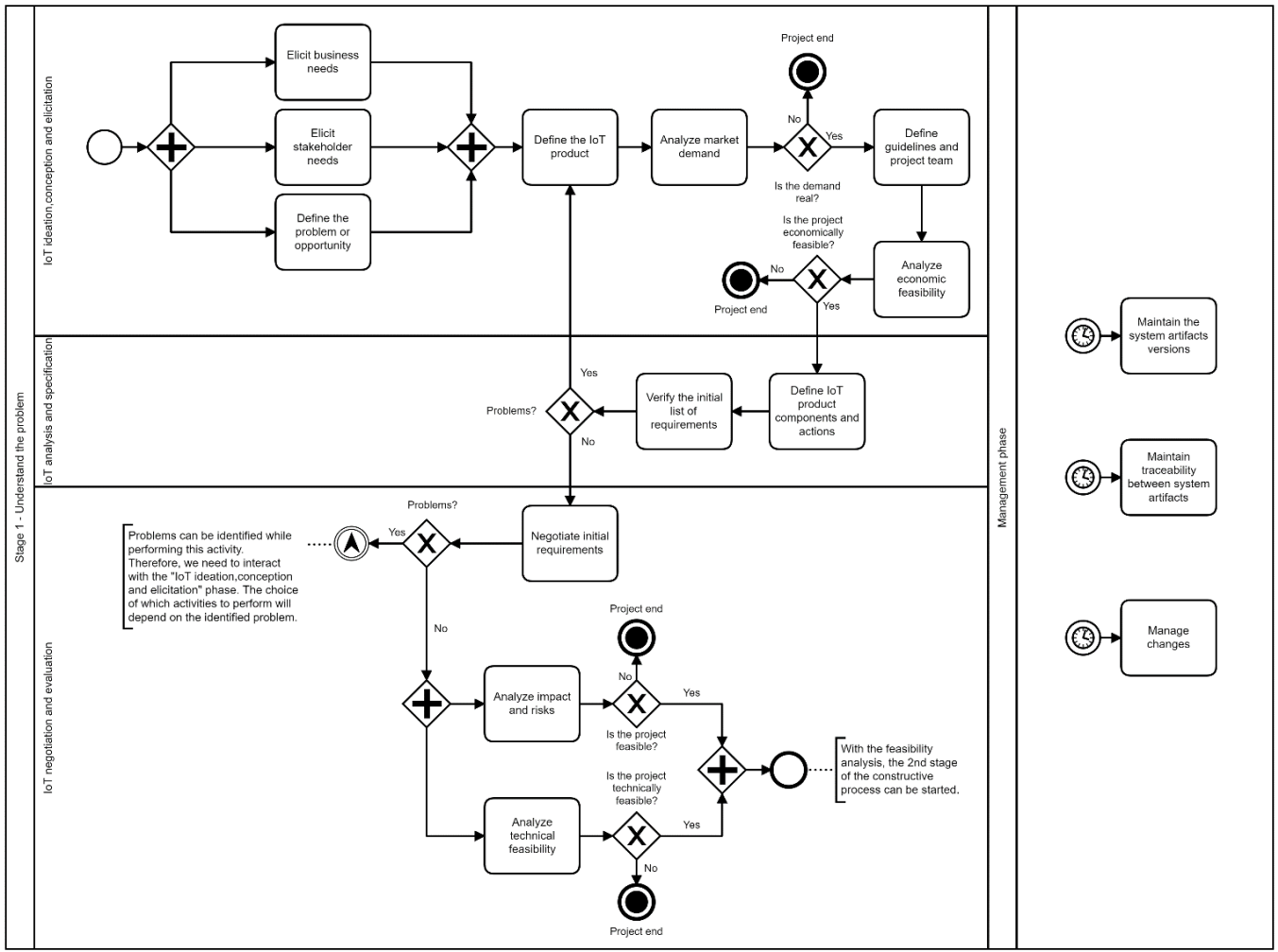

**Figure 7.** First stage - Overview of the construction process.

### 3.2.2 Stage 2

The second stage is to **describe the solution.** It aims to transform business needs, stakeholders' needs, and general requirements into detailed, classified, and organized requirements. IoT scenarios, arrangements, and components are used for the specification and verified during this stage.

Subsequently, the requirements are negotiated and evaluated, attesting that a common understanding of the system has been reached. This stage is composed of 12 activities and 39 tasks that are distributed throughout the engineering cycle.

The SCENARI$_{OT}$ technique (Silva 2019) supports the requirements identification and the system behavior description. This stage also uses the SCENARI$_{OT}$CHECK technique (Souza 2020).

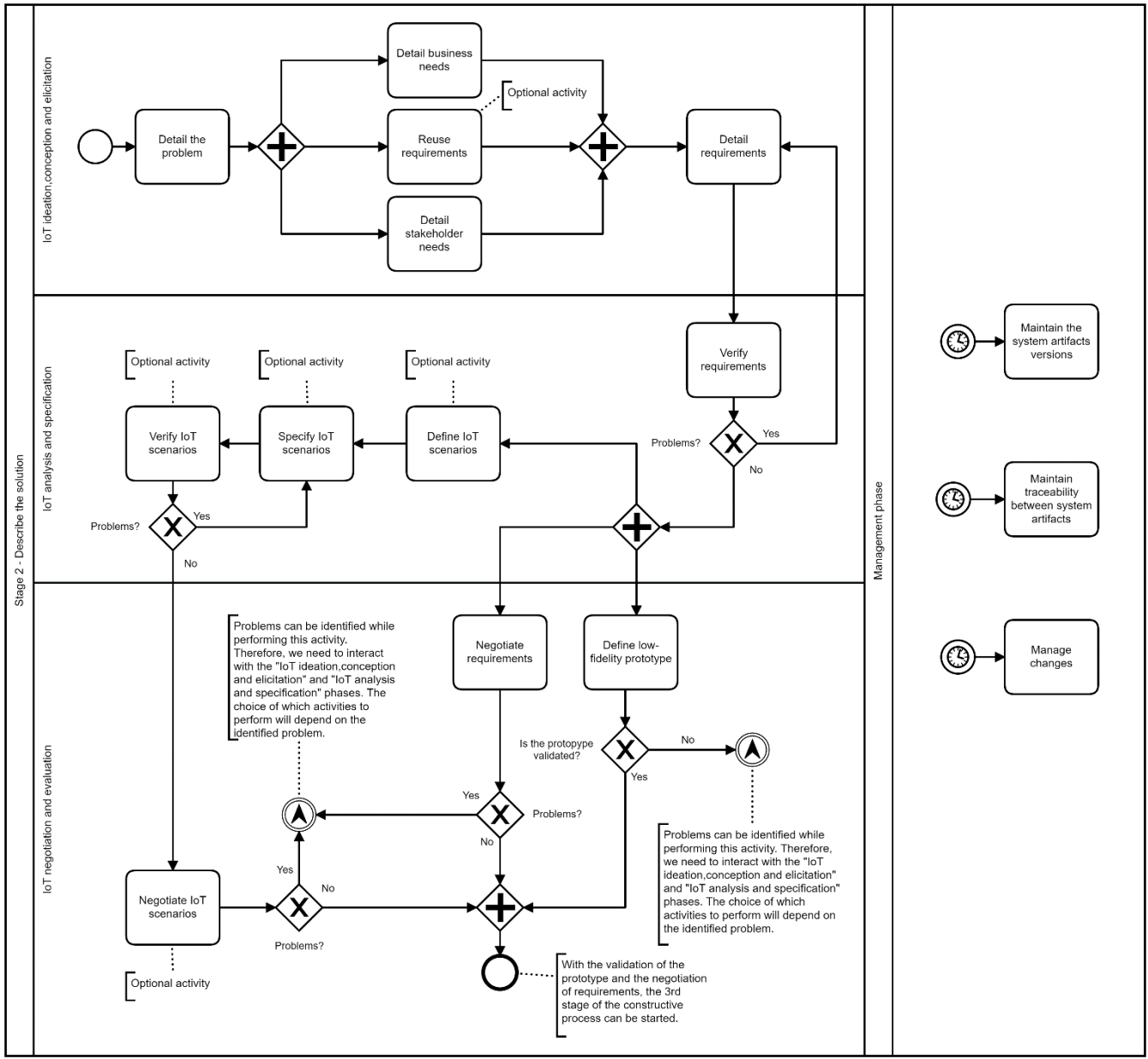

**Figure 8.** Second stage - Overview of the construction process.

**Figure 8** presents an overview of the activities performed in the second stage. This stage defines three templates: **IoT Project Detail, IoT Solution Proposal,** and **Change Analysis Report.**

The SCENARI$_{OT}$CHECK technique (Souza 2020) contributes with two templates (Verification Checklist Template and Inspection Record Template) used in this stage. Its milestone is the **Low-level Prototype** performed by the activity "Define low-fidelity prototype." This stage presents optional activities since the construction process can be used with any development methodology.

### 3.2.3 Stage 3

The third stage is to **Detail the solution.** It transforms IoT requirements and scenarios into IoT Use Cases. During this stage, the IoT Use Cases diagram, the list of IoT Use Cases, and their descriptions are generated.

Subsequently, the generated artifacts are checked and evaluated, attesting that a common understanding of the system has been achieved. This stage is composed of ten activities and 24 tasks distributed throughout the engineering cycle.

**Figure 9** presents an overview of the activities performed in the third stage. Two templates are defined for it: **IoT Use Case Description** and **IoT Diagram and Use cases Checklist**. Also, the **Change Analysis Report** can be used in this stage. This stage's milestone is the **High-level Prototype** performed by the activity "Define an evolved prototype." This stage presents optional activities since the construction process can be used with any development methodology.

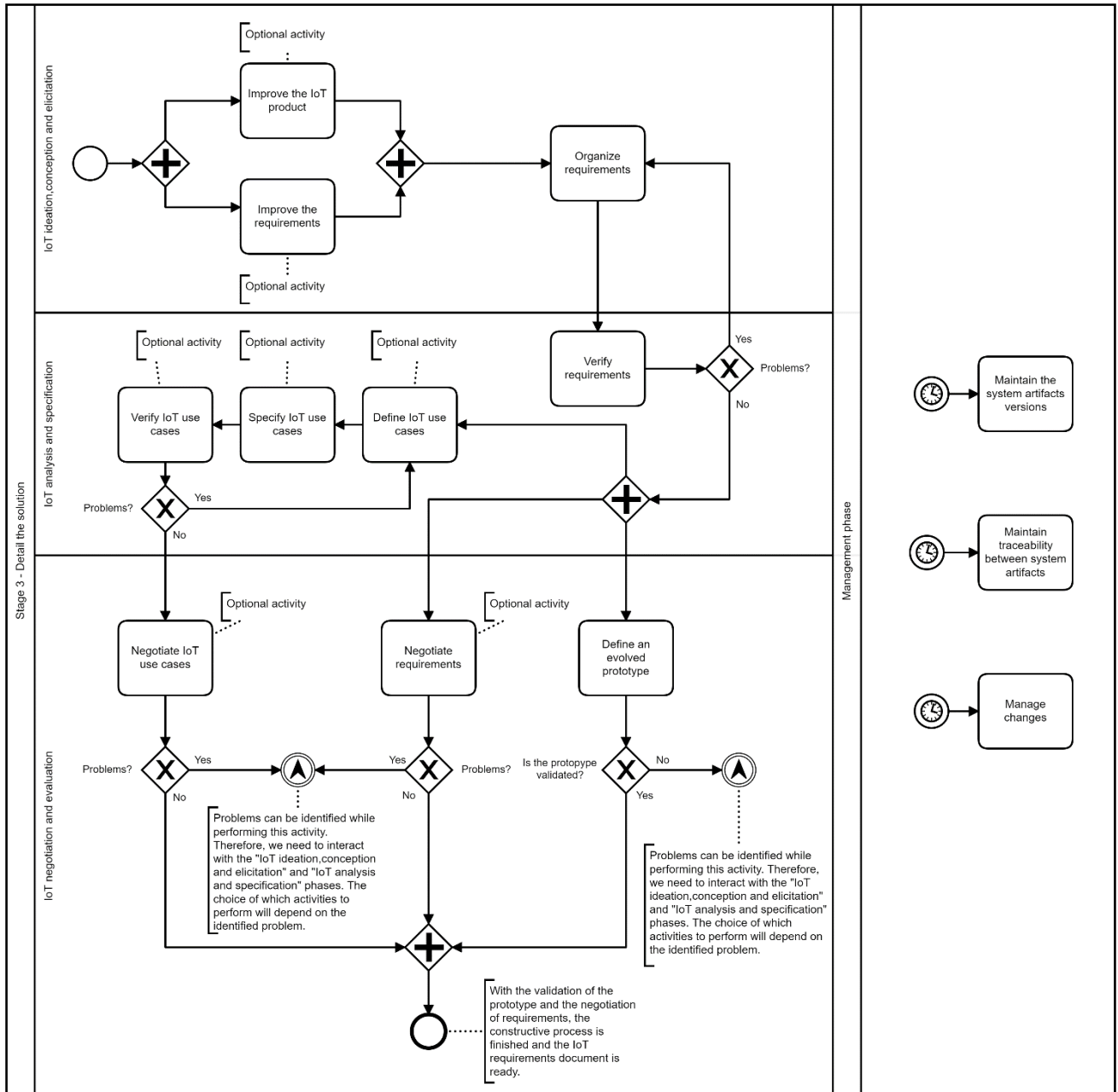

**Figure 9.** Third stage - Overview of the construction process.

# 4 Evaluating the RET$_{IoT}$'s Templates Feasibility

The RET$_{IoT}$ aims to support software engineers during the RE activities. The main techniques presented in section 2 and used to compose the software technology have already been empirically evaluated and used in IoT software system projects (Souza et al. 2019a) (Souza et al. 2019b). However, new facilities' inclusion to support the RE with the RET$_{IoT}$ requires an initial observation before using them in the projects and conducting further experimental studies. Thus, this section presents a feasibility study of the RET$_{IoT}$ templates.

## 4.1 Templates

In this feasibility study, we considered the structure of two artifacts' templates – Requirements List (RL) and IoT use-cases Description (IoTUCD1) – for conventional software systems but used in IoT software system projects. We compared their structure with the structure of the RET$_{IoT}$ templates – Project Scope (PS), Solution Proposal (SP), and IoT use-cases Description (IoTUCD2). The full versions of all templates are available at http://bit.ly/393SgHX.

### 4.1.1 The RET$_{IoT}$ Templates

This section presents three templates of the RET$_{IoT}$ (Silva et al. 2020b) regarding the activities of **elicitation** (ELI) – "Project Scope (PS)" Template (see **Figure 10**); **analysis** (ANA) and **specification** (SPE) – "Solution Proposal (SP)"

Template (see **Figure 11**) and "IoT use-cases Description (IoTUCD)" Template (see **Figure 12**). The **conception/design** (CON), **negotiation** (NEG), and **validation** (VAL) activities are minimally covered by the "Project scope" Template.

The "Solution Proposal" and "IoT use-cases Description" templates support the **management** (MAN) activities, maintaining traceability between requirements and analysis models. Besides, the techniques described in section 2 support the **elicitation** (ELI), **specification** (SPE), and **verification** (VER) activities. The following items will present the templates' global description (Project Scope, Solution Proposal, and IoT use-cases Description) as defined in the RET$_{IoT}$.

- **Project Scope Template**

This template supports the documentation of the project's initial activities, the problem to be solved, those involved in the project, the user profiles, user needs, and business needs. It includes identifying and describing system requirements (functional, non-functional, restrictions, others) and business rules.

Also, the requirements document's validation is made through an explicit agreement (signature or email copy). It provides two (status and priority) fields to support the negotiation of functional and non-functional requirements.

**Figure 10** presents an extract of this template. The proposed template is used in the activities of conception/design (CON), elicitation (ELI), negotiation (NEG), and validation (VAL).

**Figure 10.** Extract from the "Project Scope template."

- **Solution Proposal Template**

This template supports the solution description. It identifies and describes, using the SCENARI$_{OT}$ technique, the IoT scenarios, the IoT components, and the IoT interaction arrangements (IIAs) of the system. Also, it provides the details of the IIAs chosen for each IoT scenario via the corresponding catalogs. The traceability between requirements, IoT scenarios, IoT interaction arrangements, and their respective catalogs is maintained.

**Figure 11** presents an extract of this template. It should be used in the elicitation (ELI), analysis (ANA), specification (SPE), and management (MAN) activities to identify, describe and refine the system's behavior while maintaining requirements traceability.

**Figure 11.** Extract from the "Solution Proposal template."

- **IoT use-cases Description Template**

This template includes the description of the IoT use-cases. Use cases are identified and described providing a view of the system's behavior. The use-case diagram is inserted in this template.

Traceability between requirements, IoT scenarios, IoT interaction arrangements, and IoT use-cases is maintained. **Figure 12** presents an extract of this template. It should be used in the analysis (ANA), specification (SPE), verification (VER), and management (MAN) activities. The SCENARI$_{OT}$CHECK technique is applied during the verification activities to identify inconsistencies in the IoT scenarios description and their components, and the choice of IIAs.

![Figure 12 - IoT use-cases Description template extract showing a table with IoT Use Cases Description header, columns for IoT use case ID, Title, IoT Requirements, Interaction Arrangements, IoT Scenarios, followed by IoT Use Case Diagram, IoT Use Case Detail sections with fields for Use Case ID, IoT Interaction Arrangements, Preconditions, Postconditions, Associated Use Cases, Actors (Users, Things, Software systems), Interaction Sequence, and Steps (BASE FLOW, ALTERNATIVE FLOW, EXCEPTION FLOW, BUSINESS RULES).]

**Figure 12.** Extract from the "IoT use-cases Description template."

#### 4.1.2 Projects and Teams

The RL and IoTUCD1 artifacts were built by using conventional templates in three (3) IoT based software projects:

- **Project A** supports environmental markers' collection (e.g., temperature, humidity, particulates, $CO_2$ level, and toxic gases).
- **Project B** monitors a high-performance computing environment (data center) to collect different information such as temperature, environment humidity, energy consumption, and energy supply quality.
- **Project C** collects temperature, humidity, wind speed, and wind direction in different regions of a city.

All three projects represent real demand, and a stakeholder (totally external to the course and the research group) worked with the developers, including the requirements acceptance. Undergraduate students produced the RL and IoTUCD1 artifacts during a Software Engineering course at UFRJ. The course had the participation of 21 students of the fourth year of Information and Computer Engineering.

The subjects were organized into three development teams, with seven participants each. The teams contained balanced participants with equivalent levels of knowledge and skills regarding software and hardware. Training on different topics in Software Engineering and mentoring throughout the project were available. There was no intervention by the mentors in the artifact's content. All ethical issues and consent forms were made available.

Some of the course's topics included requirements engineering, IoT scenarios, verification technique for IoT systems, UML (Unified Modeling Language) diagrams, among others. The $SCENARI_{OT}$ and $SCENARI_{OT}CHECK$ techniques were presented to the participants, although they were not conditioned to use them.

The teams were free to organize their projects. The requirements document represented one of the design milestones. A minimum viable product (MVP) represents one of the concrete results delivered at the end of the course.

### 4.2 Execution

After the three teams constructed the requirements document (composed of RL and IoTUCD1 artifacts), the researchers (paper's authors) analyzed them.

The information found in the generated artifacts was compared with the requested information in the Project Scope (PS), Solution Proposal (SP), and IoT use-cases Description (IoTUCD2) templates structure. A working checklist was used to compare the templates, which will be presented in the next subsection.

Three researchers carried out the comparison – two master students and one Ph.D. that work in Software Engineering and IoT domains. After that, a fourth researcher (Ph.D. and expert in Software Engineering and IoT domains) reviewed the results' analysis.

### 4.3 Results and Discussion

**Table 1** presents the checklist used to compare the template structure (conventional software systems and $RET_{IoT}$) and the analysis result. It indicates that:

- The RL template does not address the project/system objective and problem domain. To know the problem domain is essential information for building an IoT software system (Motta et al. 2019) (Nguyen-Duc et al., 2019).
- The RL template presents a partial description of the stakeholders. It does not include profiles descriptions of the different users that are important for the system development and the user interface design.
- The RL template does not address the description of business/stakeholder needs. The identification of business/stakeholder needs represents the initial stage of the project. In this step, we seek to understand the client's real need, which will be transformed into system requirements in the future.
- $RET_{IoT}$ allows identifying the requirements that will guide the IoT solution from the beginning (Project Scope template), unlike the RL template that does not identify the IoT requirements,

**Table 1.** Mapping checklist of the template structure.

| Project/system information | Conventional templates | | RET_IoT templates | | |
|---|---|---|---|---|---|
| | RL | IoTUCD2 | PS | SP | IoTUCD2 |
| Project name/Project responsible | T | T | T | T | T |
| Version control | T | T | T | T | T |
| Explicit agreement | T | | T | | |
| Project/system objective | N | | T | | |
| Problem domain | N | | T | | |
| Project scope | T | | T | | |
| Glossaire | T | | T | | |
| Stakeholders description | P | | T | | |
| Business and Stakeholders needs a description | N | | T | | |
| Functional requirements | P | | T | | |
| Non-functional requirements | T | | T | | |
| Requirements negotiation (prioritization) | T | | T | | |
| Business rules | N | T | T | | T |
| Project analyses | N | | P | | |
| IoT scenarios | N | P | | T | T |
| IoT components description | | N | | T | T |
| IoT interaction arrangements | | P | | T | T |
| IoT use-cases diagram | | T | | | T |
| IoT use-cases description | | T | | | T |
| Traceability | | P | | T | T |
| References (others project documents) | T | | N | | |

P - Partially collected; T - Totally collected; N - Does not collect information; Gray - Not applicable for this template.

- The IoTUCD1 template treats IoT scenarios and IoT arrangements partially but does not address IoT components' description. In contrast, the RET_IoT treats this information entirely in the Solution Proposal (SP) and IoT use-cases Description (IoTUCD 2) templates.
- The traceability is partially treated by the IoTUCD1 template and fully treated by RET_IoT in two templates (SP and IoTUCD2).
- The RL template presents a field for references (other documents), which RET_IoT does not address.

The different convergence and divergence points between conventional systems templates (RL and IoTUCD1) and RET_IoT templates (PS, SP, IoTUCD2) offer indications that the RET_IoT can be more robust because it deals with IoT information since the beginning of the project.

According to the results, RET_IoT can present a good potential for supporting IoT software systems' specifications because of its templates' specific IoT information, differently from conventional ones.

### 4.4 Templates' evolution

This study allowed us to evolve the existing templates regarding the reorganization of sections and the insertion of new sections and new fields. The **Project Scope** template was renamed to **IoT Project Detail** and the **Solution Proposal** template to **IoT Solution Proposal**.

In the **IoT Project Detail** template, we included a new field, "Project description." The "Glossary" and "Stakeholders" sections have been changed to include fields to support the capturing of specific information. The "Potential stakeholders" section has been changed to "Stakeholders" to include two new fields to capture each stakeholder's interest and its influence in the system. The "Project scope" section has been removed from the template. The "Canvas IoT" section was added, allowing the insertion of an image or photo of the IoT Canvas built in Stage 1.

New fields ("Reused requirement?" and "Related requirement ID") have been added to the "System requirements" section to enable requirements traceability (functional and non-functional). For functional requirements, two fields ("Cost" and "Effort") were added to make negotiation feasible. In its previous version, requirements were classified into IoT requirements and non-IoT requirements. In this new version, this classification has been removed, and the "IoT Characteristic" field has been included. Therefore, when describing a non-IoT requirement, this field should not be filled. In an IoT requirement, the IoT characteristic must be described as identification, sensing, performance, connectivity, and processing. The "Dependency between requirements" field has also been added to the "Functional requirements" section.

The non-functional requirement "scalability" was added to the new template. The requirements "portability and compatibility" and "security and privacy" have been adapted.

The section "Annex - Non-functional requirements" has been added to support the identification of non-functional requirements. The section "Scope not covered by the project" has been added, and the section "Project analysis" has been removed. In the "Business rules" section, the "Related needs ID" field has been added to allow business rules' traceability.

In the **IoT Solution Proposal** template, the fields "Actors," "Actions," "Interaction Arrangements" were added in the "IoT Scenarios" section. The section "IoT system components" was removed because it had a redundancy of the arrangement catalogs' information. The "Related functional requirements," "Precedencies," and "Dependencies" fields have been added in the "IoT scenarios description" section to enable the traceability of IoT scenarios. The field "Collected data and Actions performed" was divided into two fields: "Collected data" and "Actions performed." The "Interaction sequence" field was changed because it is like a use-case structure (main, alternative, and exception flows). The "Environment" and "Connectivity" fields have been removed.

In the **IoT use-cases Description** template, the "Business rules" field was moved from the "Interaction sequence" section to a separate section. The section "Customer or customer representative agreement" has been added to this template.

Five new templates were also defined to support the construction process activities that had not yet been contemplated. The new models, cited in section 3, correspond to **IoT Canvas, IoT Project Feasibility Analysis, Requirements Checklist, Change Analysis Report,** and **IoT Diagram and Use cases Checklist**.

However, to ensure the technology validity, further experimental evaluation is necessary to verify whether the $RET_{IoT}$ construction process with the templates is useful, complete, correct, and intuitive.

### 4.5 Threats to Validity

I**nternal validity** is the study itself, even though experimental studies have evaluated part of the RETIoT technology. However, the results indicate that the $RET_{IoT}$ templates can capture relevant information compared to conventional templates regarding project artifacts.

An **external validity** issue concerns the participants (undergraduate students) who have been invited to participate in the study. We cannot claim that the information provided is complete from the project's point of view, nor did the participants understand all the topics taught during the course. To mitigate this threat, the projects treated in the study represented real problems. Besides, each team had contact with a stakeholder of each addressed problem.

There was no control over the artifact's creation during the course and used in the study regarding their construct validity. The projects were equivalent in size, complexity and used IoT technologies to mitigate this threat. Also, it can be highlighted that the teams received equivalent training and mentoring in RE.

Finally, the **conclusion validity** concerns the study interpretation and sample size. We had a small and inhomogeneous sample size. Therefore, it was impossible to apply statistical tests to carry out a deeper analysis of the results obtained. Also, the study conclusion is limited to the researchers' interpretation. These items limit the study results generalization. To mitigate this threat, we aim to perform future experimental studies to collect feedback from the $RET_{IoT}$.

## 5 Related Work

This section presents a set of related works found in the technical literature, which address technologies for the different RE phases mentioned above. **Table 2** presents a comparison of seventeen (17) technologies found in the technical literature. We can observe that conception, negotiation, verification, validation, and management phases need more attention regarding IoT concepts and characteristics.

**Figure 13** synthesizes the information presented in **Table 2**, showing the number of technologies per RE phase. We can highlight that a high number permeate elicitation (nine), analysis (ten), and specification (eight) phases, while a small number is concentrated in the conception/design (four), negotiation (one), verification (five), validation (three) and management (three) phases.

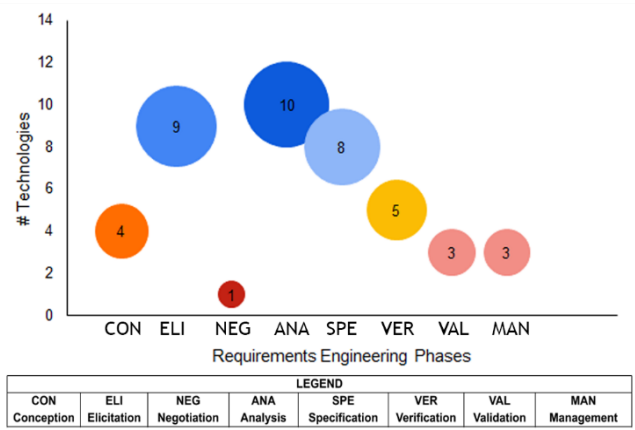

**Figure 13.** Technologies x RE phases.

Regarding the **conception phase** (CON), GSEM-IoT (Zambonelli 2017) (Laplante et al. 2018) and Ignite (Giray et al. 2018) technologies carry out the stakeholders' analysis involved in the system. The feasibility analysis is partially addressed by IoT Methodology (Giray et al., 2018). Also, the Ignite and CORE (Hamdi et al. 2019) technologies provide business analysis mechanisms.

Several technologies address the elicitation phase (ELI): Ignite (Giray et al. 2018), IoT Methodology (Giray et al. 2018), (Laplante et al. 2018), IoTReq (Reggio 2018), (Curumsing et al. 2019), CORE (Hamdi et al. 2019), SCENARI$_{OT}$ (Silva 2019) and TrUStAPIS (Ferraris and Fernandez-Gago 2020) that offer resources for collecting requirements. GSEM-IoT (Zambonelli 2017), IoTReq, and IoT Methodology propose mechanisms to transform users' needs into requirements.

**Table 2.** Technologies x RE phases.

| Technology/RE Phase | CON | ELI | NEG | ANA | SPE | VER | VAL | MAN |
|---|---|---|---|---|---|---|---|---|
| (Aziz et al. 2016) | | | | x | x | | | x |
| (Mahalank et al. 2016) | | | | | x | | | |
| (Takeda and Hatakeyama 2016) | | | | x | x | | | |
| (Touzani and Ponsard 2016) | | | | x | | | | |
| IoT-RML (Costa et al. 2017) | | | | x | x | x | | |
| (Yamakami 2017) | | | | | | x | | |
| GSEM-IoT (Zambonelli 2017) | x | x | | | | | | |
| (Carvalho et al. 2018) | | | | | | x | | |
| (Curumsing et al. 2019) | | x | | x | | x | | x |
| IoT System Development Methods — Ignite (Giray et al. 2018) | x | x | x | x | x | | x | |
| IoT System Development Methods — IoT Methodology (Giray et al. 2018) | x | x | | | | | x | |
| (Laplante et al. 2018) | x | x | | x | | | x | |
| IoTReq (Reggio 2018) | | x | | x | x | | | |
| CORE (Hamdi et al. 2019) | | x | | x | | | | |
| SCENARI_OT (Silva 2019) | | x | | x | x | | | |
| SCENARI_OT CHECK (Souza 2020) | | | | | | x | | |
| TrUStAPIS (Ferraris and Fernandez-Gago 2020) | | x | | | x | | | x |

For the **negotiation phase** (NEG), Ignite (Giray et al. 2018) addresses the impact and risk analysis but does not provide further details on conducting this activity.

In the **analysis phase** (ANA), (Takeda and Hatakeyama 2016) and (Touzani and Ponsard 2016) technologies, Ignite (Giray et al. 2018), (Laplante et al. 2018), IoTReq (Reggio 2018), (Curumsing et al. 2019) and CORE (Hamdi et al. 2019) use UML diagrams to develop the analysis models. The SCENARI_OT technology (Silva 2019) comprises the scenario analysis based on IoT interaction arrangements. The works of (Aziz et al. 2016) and IoT-RML (Costa et al., 2017) address artifacts and models' reuse.

The **specification phase** (SPE) is addressed by several technologies: (Takeda and Hatakeyama 2016), IoT-RML (Costa et al. 2017), IoTReq (Reggio 2018), and TrUStAPIS (Ferraris and Fernandez-Gago 2020) that use formal models for specifying requirements. Technologies proposed by (Aziz et al. 2016), (Mahalank et al. 2016), and (Giray et al. 2018) – Ignite provide templates for specifying requirements. The SCENARI_OT (Silva 2019) proposes the scenario specification using IoT interaction arrangements.

In the **verification phase** (VER), we found that (Carvalho et al. 2018) and SCENARI_OT CHECK (Souza 2020) propose mechanisms to verify requirements. The technologies proposed by (Yamakami 2017), (Costa et al. 2017) - IoT-RML (Carvalho et al., 2018), and (Curumsing et al. 2019) offer mechanisms for checking conflicts between requirements.

The **validation phase** (VAL) is addressed by Ignite (Giray et al., 2018), IoT Methodology (Giray et al., 2018), and (Laplante et al., 2018), which propose a prototyping technique to ensure that the product meets users' needs.

For the **management phase** (GER), (Aziz et al. 2016), (Curumsing et al. 2019), and TrUStAPIS (Ferraris and Fernandez-Gago 2020) offer mechanisms to enable traceability. TrUStAPIS also provides a mechanism for requirements change management.

A quasi- systematic literature review (Lim et al. 2018) identified 12 relevant publications and 37 elicitation techniques normally applied in IoT systems development. The most frequently used techniques are **interviews** and **prototypes**, where the latter can also be used to validate requirements.

We can also highlight other techniques and methods applied during the elicitation phase: scenarios, use cases, and frameworks. This work also presents a brief contribution regarding the conflict resolution of the stakeholders. The authors emphasize using interview and prototyping techniques to encourage discussions and find alternative ways to the identified conflicts.

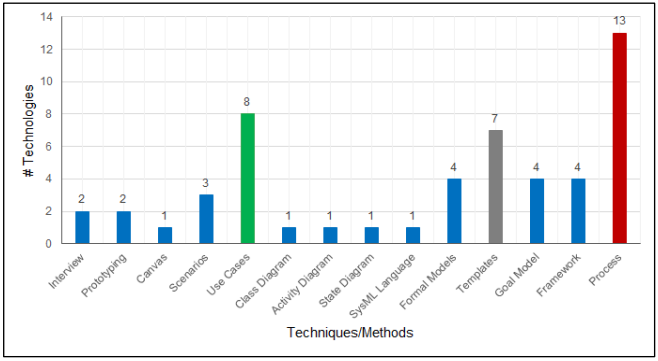

**Figure 14.** Technologies x Techniques/Methods.

**Table 3.** Techniques/Methods x RE phases.

| Techniques-Methods /RE Phase | CON | ELI | NEG | ANA | SPE | VER | VAL | MAN |
|---|---|---|---|---|---|---|---|---|
| Interview | | 2 | | | | | | |
| Prototyping | | | | | | | 3 | |
| Canvas | | 1 | | | | | 1 | |
| Scenarios | | 1 | | | 3 | | | |
| Use cases | 2 | 2 | | 7 | 1 | | | |
| Class Diagram | | | | 1 | | | | |
| Activity Diagram | | | | 1 | | | | |
| State Diagram | | | | 1 | | | | |
| SysML Language | | | | | 1 | 1 | 1 | |
| Formal Models | | 2 | | 4 | 2 | 1 | | 1 |
| Templates | 2 | 3 | | | 5 | 1 | | 2 |
| Goal model | | 2 | | 4 | 1 | | | 1 |
| Framework | 2 | 3 | 1 | 2 | 1 | | 2 | |
| Catalogs | | 1 | | | 1 | 1 | | |
| Process | 3 | 11 | 1 | 9 | 7 | 3 | 3 | 3 |
| Total | 9 | 28 | 2 | 30 | 22 | 7 | 9 | 7 |

In this way, we analyzed the 17 technologies to identify which techniques/methods are used and where (RE phases) in IoT systems development. **Figure 14** shows our findings where we can observe 14 items and the most used: process (thirteen), use cases (eight), and models (seven).

**Table 3** shows where (RE phases) the techniques/methods found are applied. The elicitation (28), analysis (30), and specification (22) phases offer a greater number of techniques/methods. It is important to highlight that some technologies offer more than one technique for one or more RE phases.

RET$_{IoT}$ permeates the eight phases previously described offering methodological and technical support through construction, techniques, and templates.

Analyzing the RET$_{IoT}$ current version (see section 3), we can say that it proposes and integrates some techniques/methods: prototyping, IoT canvas, IoT scenarios based on IoT scenarios specification technique - SCENAR-I$_{OT}$ (Silva 2019), use cases diagram and description, templates, and IoT scenario inspection technique - SCENARI-$_{OT}$CHECK (Souza 2020), and a construction process.

## 6 Research Opportunities

Analyzing the technologies found in the technical literature, we can observe that only one technology discusses the **Negotiation phase**. It represents a research opportunity. Few technologies offer project management, validation, test case elaboration, and decision-making related to the system's design and architecture. These topics can be explored through future research. We can also observe that not all technologies cover all RE activities and present gaps regarding the different activities necessary to build IoT system requirements documents.

Among these gaps, we can observe the lack of i) methodological support for the design and ideation of IoT products (Nguyen-Duc et al. 2019); ii) stakeholder identification and description and business needs (Silva et al. 2020b); iii) IoT system characteristics and behaviors (Motta et al. 2019a), as well as the requirements refinement; iv) high-level (new IoT interaction arrangements) and low-level (IoT use-case diagram) analysis models; v) project feasibility analysis (Silva et al. 2020); vi) prototypes as suggested by (Nguyen-Duc et al. 2019) (Lim et al. 2018); and vii) explicit agreements with the client (Silva et al. 2020).

These technologies also do not fully meet the IoT software system specificities and characteristics: i) the components and actors' description (Curumsing et al. 2019) (Aziz et al. 2016); ii) the behaviors description of different levels of each object - (Curumsing et al. 2019) (Reggio 2018); iii) the identification of the systemic characteristics (sensing, identification, performance, processing, and connectivity); and iv) the detailed specification of each feature.

# 7 Conclusion and Future Works

This paper presented the RET$_{IoT}$, which aims to provide a construction process, techniques (IoT scenario specification and verification techniques), and tools (templates) to support IoT software's Requirements Engineering systems. A feasibility study was performed to compare three templates defined in the RET$_{IoT}$ with conventional software systems templates (not specific to IoT software systems). Their comparison provided indications that the artifacts generated by RET$_{IoT}$ may be complete regarding the capture of IoT information.

Some of the future works reserved for the RET$_{IoT}$ are: i) design and execution of experimental studies to evaluate the technology in more robust IoT software system projects (academic and industrial contexts); A comparative study of the RET$_{IoT}$ with traditional technologies will be carried out to verify the efficiency and effectiveness of the RET$_{IoT}$ in terms of capturing system and project relevant information. The study will also evaluate the RET$_{IoT}$ usefulness and suitability according to the user's perception; ii) the integration of CATS# (Context-Aware Test Suite Design) technique (Doreste and Travassos 2020) with the RET$_{IoT}$. CATS# is a testing technique to support software engineers with the specification of test cases, capturing the context and variations, and iii) developing tooling support that integrates the construction process, IoT scenario specification, and verification techniques templates. The tool will facilitate the traceability among IoT requirements, IoT Interaction Arrangements, IoT scenarios, and IoT use-cases.

# Acknowledgments

The authors would like to thank the National Council for Scientific and Technological Development - CNPq. Taisa Gonçalves received a postdoctoral scholarship (154004 / 2018-9). Prof. Travassos is a CNPq researcher (304234/2018-4).# References

Alexander I, Maiden N (2004) Scenarios, stories, and use cases: the modern basis for system development. Computing and Control Engineering 15:24–29. https://doi.org/10.1049/cce:20040505

Arif S, Khan Q, Gahyyur SAK (2009) Requirements engineering processes, tools/technologies, & methodologies. International Journal of Reviews in Computing 2:41–56

Aziz MW, Sheikh AA, Felemban EA (2016) Requirement Engineering Technique for Smart Spaces. In: International Conference on Internet of things and Cloud Computing. ACM Press, Cambridge, United Kingdom, p 54:1-54:7

Behrens H (2002) Requirements Analysis Using Statecharts and Generated Scenarios. In: Doctoral Symposium at IEEE Joint Conference on Requirements Engineering. IEEE, Essen, Germany, pp 1–5

Burg JFM, Van de Riet RP (1996) A Natural Language and Scenario based Approach to Requirements Engineering. In: Proceedings of Workshop in Natuerlichsprachlicher Entwurf von Informationssystemen. Tutzing, Germany, pp 219–233

Carvalho RM, Andrade RMC, Oliveira KM (2018) Towards a catalog of conflicts for HCI quality characteristics in UbiComp and IoT applications: Process and first results. In: 12th International Conference on Research Challenges in Information Science (RCIS). IEEE, Nantes, pp 1–6

Costa B, Pires PF, Delicato FC (2017) Specifying Functional Requirements and QoS Parameters for IoT Systems. In: 15th Intl Conf on Dependable, Autonomic and Secure Computing, 15th Intl Conf on Pervasive Intelligence and Computing, 3rd Intl Conf on Big Data Intelligence and Computing and Cyber Science and Technology Congress. IEEE, Orlando, United States, pp 407–414

Curumsing MK, Fernando N, Abdelrazek M, et al. (2019) Emotion-oriented requirements engineering: A case study in developing a smart home system for the elderly. Journal of Systems and Software 147:215–229. https://doi.org/10.1016/j.jss.2018.06.077

Doreste AC de S, Travassos GH (2020) Towards Supporting the Specification of Context-Aware Software System Test Cases. In: XXIII Ibero-American Conference on Software Engineering. Springer, Curitiba, Brazil (Online), p S10 P1:8 pages

Ferraris D, Fernandez-Gago C (2020) TrUStAPIS: a trust requirements elicitation method for IoT. International Journal of Information Security 19:111–127. https://doi.org/10.1007/s10207-019-00438-x

Giray G, Tekinerdogan B, Tüzün E (2018) IoT System Development Methods. In: Hassan Q, Khan AR, Madani SA (eds) Internet of Things, 1ª Edição. CRC Press/Taylor & Francis, New York, pp 141–159

Glinz M (2000) Improving the Quality of Requirements with Scenarios. In: Proceedings of the World Congress on Software Quality (WCSQ). Yokohama, Japan, pp 55–60

Hamdi MS, Ghannem A, Loucopoulos P, et al. (2019) Intelligent Parking Management by Means of Capability Oriented Requirements Engineering. In: Wotawa F, Friedrich G, Pill I, et al. (eds) Advances and Trends in Artificial Intelligence - From Theory to Practice - IEA/AIE 2019. Springer International Publishing, Cham, pp 158–172

Laplante NL, Laplante PA, Voas JM (2018) Stakeholder Identification and Use Case Representation for Internet-of-Things Applications in Healthcare. IEEE Systems Journal 12:1589–1597. https://doi.org/10.1109/JSYST.2016.2558449

Lim T-Y, Chua F-F, Tajuddin BB (2018) Elicitation Techniques for Internet of Things Applications Requirements:

A Systematic Review. In: VII International Conference on Network, Communication, and Computing. ACM Press, Taipei City, Taiwan, pp 182–188

Mahalank SN, Malagund KB, Banakar RMB (2016) Non-Functional Requirement Analysis in IoT based smart traffic management system. In: International Conference on Computing Communication Control and Automation. IEEE, Pune, India, pp 1–6

Motta RC, Oliveira KM de, Travassos GH (2019a) On challenges in engineering IoT software systems. Journal of Software Engineering Research and Development 7:5:1-5:20. https://doi.org/10.5753/jserd.2019.15

Motta RC, Oliveira KM, Travassos GH (2020) Towards a Roadmap for the Internet of Things Software Systems Engineering. In: Proceedings of the 12th International Conference on Management of Digital EcoSystems. ACM, Virtual Event, United Arab Emirates, pp 111–114

Motta RC, Silva VM, Travassos GH (2019b) Towards a more in-depth understanding of the IoT Paradigm and its challenges. Journal of Software Engineering Research and Development 7:3:1-3:16. https://doi.org/10.5753/jserd.2019.14

Nguyen-Duc A, Khalid K, Shahid Bajwa S, Lønnestad T (2019) Minimum Viable Products for the Internet of Things Applications: Common Pitfalls and Practices. Future Internet 11:50:1-50:21. https://doi.org/10.3390/fi11020050

Pandey D, Suman U, Ramani AK (2010) An Effective Requirement Engineering Process Model for Software Development and Requirements Management. In: International Conference on Advances in Recent Technologies in Communication and Computing. IEEE, Kottayam, India, pp 287–291

Pressman RS (2014) Software engineering: a practitioner's approach, 8th ed. McGraw-Hill Higher Education, United States

Reggio G (2018) A UML-based proposal for IoT system requirements specification. In: 10th International Workshop on Modelling in Software Engineering. ACM Press, Gothenburg, Sweden, pp 9–16

Silva DV da, Gonçalves TG, Rocha ARC da (2019) A Requirements Engineering Process for IoT Systems. In: XVIII Brazilian Symposium on Software Quality. ACM Press, Fortaleza, Brazil, pp 204–209

Silva DV da, Gonçalves TG, Travassos GH (2020a) A Technology to Support the Building of Requirements Documents for IoT Software Systems. In: XIX Brazilian Symposium on Software Quality. ACM Press, São Luís, Brazil (Online), p 10 pages (to appear)

Silva DV da, Souza BP de, Gonçalves TG, Travassos GH (2020b) Uma Tecnologia para Apoiar a Engenharia de Requisitos de Sistemas de Software IoT. In: XXIII Ibero-American Conference on Software Engineering. Springer, Curitiba, Brazil (Online), p S09 P3:14 pages

Silva VM (2019) SCENARIoT Support for Scenario Specification of Internet of Things-Based Software Systems. Master's Dissertation, Federal University of Rio de Janeiro

Sommerville I (2015) Software Engineering, 10 edition. Pearson, Harlow

Souza BP de (2020) SCENARIOTCHECK: Uma Técnica de Leitura Baseada em Checklist para Verificação de Cenários IoT. Master's Dissertation, Federal University of Rio de Janeiro

Souza BP de, Motta RC, Costa D de O, Travassos GH (2019a) An IoT-based Scenario Description Inspection Technique. In: XVIII Brazilian Symposium on Software Quality. ACM Press, Fortaleza, Brazil, pp 20–29

Souza BP de, Motta RC, Travassos GH (2019b) The first version of SCENARIotCHECK: A Checklist for IoT based Scenarios. In: XXXIII Brazilian Symposium on Software Engineering. ACM Press, Salvador, Brazil, pp 219–223

Takeda A, Hatakeyama Y (2016) Conversion Method for User Experience Design Information and Software Requirement Specification. In: Marcus A (ed) Design, User Experience, and Usability: Design Thinking and Methods - DUXU 2016. Springer, Cham, pp 356–364

Touzani M, Ponsard C (2016) Towards Modelling and Analysis of Spatial and Temporal Requirements. In: 2016 IEEE 24th International Requirements Engineering Conference (RE). IEEE, Beijing, China, pp 389–394

Vegendla A, Duc AN, Gao S, Sindre G (2018) A Systematic Mapping Study on Requirements Engineering in Software Ecosystems. Journal of Information Technology Research 11:4:1-4:21. https://doi.org/10.4018/JITR.2018010104

Yamakami T (2017) Horizontal Requirement Engineering in Integration of Multiple IoT Use Cases of City Platform as a Service. In: 2017 IEEE International Conference on Computer and Information Technology (CIT). IEEE, Helsinki, Finland, pp 292–296

Zambonelli F (2017) Key Abstractions for IoT-Oriented Software Engineering. IEEE Software 34:38–45. https://doi.org/10.1109/MS.2017.3